\documentclass[12pt]{iopart}
\usepackage{graphicx}
\usepackage{amssymb}

\begin{document}

\title{The effect of Cr impurity to superconductivity in electron-doped BaFe$_{2-x}$Ni$_{x}$As$_{2}$}

\author{Rui Zhang$^{1}$, Dongliang Gong$^1$, Xingye Lu$^{1}$, Shiliang Li$^{1,3}$, Pengcheng Dai$^{2,1}$, and Huiqian Luo$^1$$^{*}$}

\address{$^1$ Beijing National Laboratory for Condensed Matter
Physics, Institute of Physics, Chinese Academy of Sciences, Beijing 100190, China}
\address{$^2$ Department of Physics and Astronomy, Rice University, Houston, Texas 77005, USA}
\address{$^3$ Collaborative Innovation Center of Quantum Matter, Beijing, China}
 \ead{$^{*}$hqluo@iphy.ac.cn}

\begin{abstract}
We use transport and magnetization measurements to study the effect of Cr-doping to the phase diagram of the electron-doped superconducting
 BaFe$_{2-x}$Ni$_{x}$As$_{2}$ iron pnictides.  In principle, adding Cr to electron-doped BaFe$_{2-x}$Ni$_{x}$As$_{2}$ should be equivalent to the effect of hole-doping.  However, we find that Cr doping suppresses superconductivity via impurity effect, while not affecting the normal state resistivity above 100 K.
We establish the phase diagram of Cr-doped BaFe$_{2-x-y}$Ni$_{x}$Cr$_y$As$_{2}$ iron pnictides,
and demonstrate that Cr-doping near optimal superconductivity restore the long-range antiferromagnetic order suppressed by superconductivity.
\end{abstract}

\section{Introduction}

A determination of the impurity effect on high-temperature (high-$T_c$) superconductors is important for establishing the microscopic origin of superconductivity in these materials \cite{Balatsky}.  For high-$T_c$ copper oxide superconductors, nonmagnetic impurities such as Zn doped into the CuO$_2$ plane dramatically suppress superconductivity and induce antiferromagnetic (AF) excitations in hole-doped superconductors \cite{Suchaneck}.
In the case of iron pnictides, electron-doping into the parent compounds can simultaneously suppress antiferromagnetic (AF) order and induce superconductivity \cite{Hosono,Cruz,JZhao}.
For the electron doped BaFe$_{2-x}$Ni$_{x}$As$_{2}$, extensive transport, x-ray and neutron diffraction measurements have established the overall structural and magnetic phase diagram \cite{ASSefat,QHuang,NNi,YCChen,HQLuo1,XYLu1,XYLu2}.
For samples near optimal superconductivity, an incommensurate AF short-range ordered phase was found to coexist and compete with superconductivity \cite{HQLuo1,DKPratt}.  Instead of originating from a spin-density-wave order from nested Fermi surfaces between the hole and electron pockets \cite{DKPratt}, the incommensurate AF short-range order was found to be consistent with a cluster spin glass mesoscopically coexisting with superconductivity \cite{XYLu2}.  If this is indeed the case, it would be interesting to determine the ground state of the system when superconductivity is eliminated, as this could potentially reveal the presence of an avoided quantum critical point in the system \cite{XYLu1}.

One way to suppress superconductivity is by applying a magnetic field, and our previous elastic neutron scattering experiment indeed show that a field suppresses superconductivity also enhances the incommensurate AF order, confirming the competing nature of the incommensurate AF phase with superconductivity \cite{HQLuo1}.  However, the maximum field one can apply in a scattering experiment is well below the upper critical field $H_{c2}$ in iron pnictide superconductors \cite{HQYuan} and an external magnetic field may also
 induce new electronic phases not present in the zero field \cite{SESebastian}.
Alternatively, superconductivity may be suppressed by the impurity substitutions in the conductance plane.
In the case of copper oxide superconductors, a small amount of impurity such as Zn, Ni, Mn, and Fe
in the CuO$_2$ plane can completely eliminate superconductivity, leading to new electronic states \cite{GXiao,JLTallon,HAlloul}.
For superconducting iron pnictides, previous studies on  the impurity effects mainly focus on the Zn, Mn, Cu, Ni and Ru substitution of Fe \cite{PCheng1,YFGuo,JLi1,JLi2,JLi3,PCheng2,YKLi,TInabe}. However, the $T_c$ suppression rate (reduction of $T_c$ for per percent concentration of impurity) of most impurities is lower than that of the cuprates, leading to the conclusion that superconducting electron pairing is inconsistent with quasiparticle excitations from the nested hole and electron Fermi surfaces \cite{SOnari,CTarantinii,YLi}.  Compared with other impurities, substituting Mn into optimally doped superconductor Ba$_{0.5}$K$_{0.5}$Fe$_2$As$_2$ has the largest superconducting suppression effect at about 7 K per 1\% Mn substitution \cite{JLi2}.

In this article, we report transport and bulk magnetic measurements on
the Cr-doped superconducting BaFe$_{2-x}$Ni$_{x}$As$_{2}$.  By carrying out systematic transport measurements on
 BaFe$_{2-x-y}$Ni$_{x}$Cr$_{y}$As$_{2}$ single crystals with different Ni and Cr concentrations,
we are able to map out the evolution of superconducting dome as a function of the Ni doping and Cr concentration (Fig. 1).
We find that optimal superconductivity in BaFe$_{2-x}$Ni$_{x}$As$_{2}$ with maximum $T_c=$ 20 K can be completely suppressed
by 1.5\% Cr doping ($y$=0.03), leading to a $T_c$ suppression rate of 13 K per 1\% Cr-doping, much
larger than that of Mn-doped iron pnictide \cite{JLi2}. From the temperature dependence of the resistivity measurements
on BaFe$_{2-x-y}$Ni$_{x}$Cr$_{y}$As$_{2}$, we find that Cr doping does not affect $T_N$, $T_s$, and
the normal state resistivity above 100 K in the
underdoped compounds $x=$ 0.049 and 0.066 until superconductivity is completely suppressed.
In the optimal doped sample $x=$ 0.082, $T_N$ firstly decreases with increasing Cr doping and then increases upon further Cr substitution.
For the electron overdoped sample with $x=$ 0.098, increasing Cr doping to $y\geq 0.033$ may induce AF order. Our results suggest that
Cr doping is effective in suppressing superconductivity, and the transport measurements provide the basis for future
neutron scattering experiments detailing the relationship between magnetism and superconductivity.

\section{Experiment}

\begin{figure}[t]
\center\includegraphics[scale=.5]{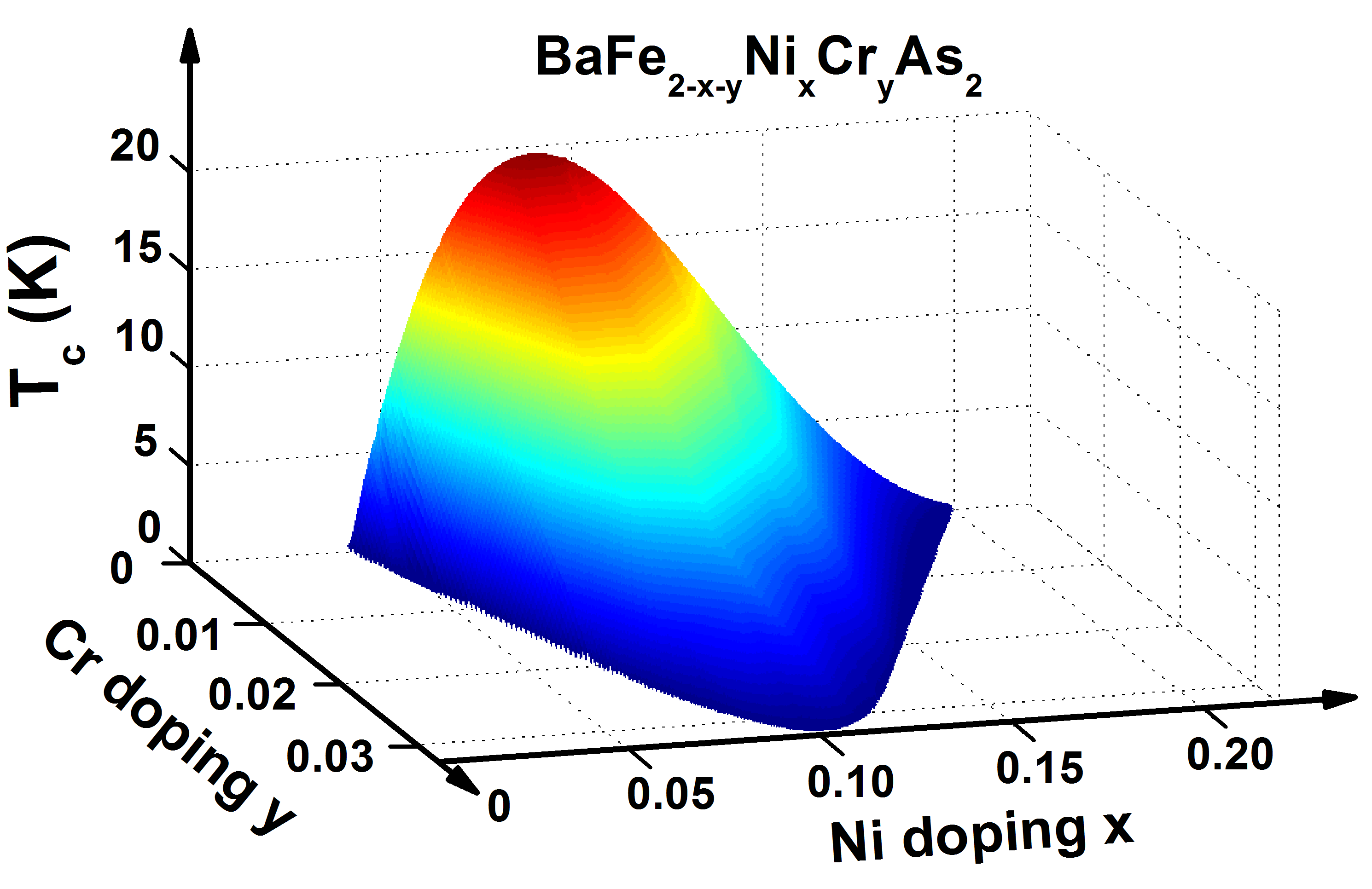}
\caption{
Three dimensional superconducting dome for BaFe$_{2-x-y}$Ni$_{x}$Cr$_{y}$As$_{2}$ as a function of Ni doping $x$ and Cr impurity $y$.
}
\end{figure}

 \begin{figure}[t]
\center\includegraphics[scale=.4]{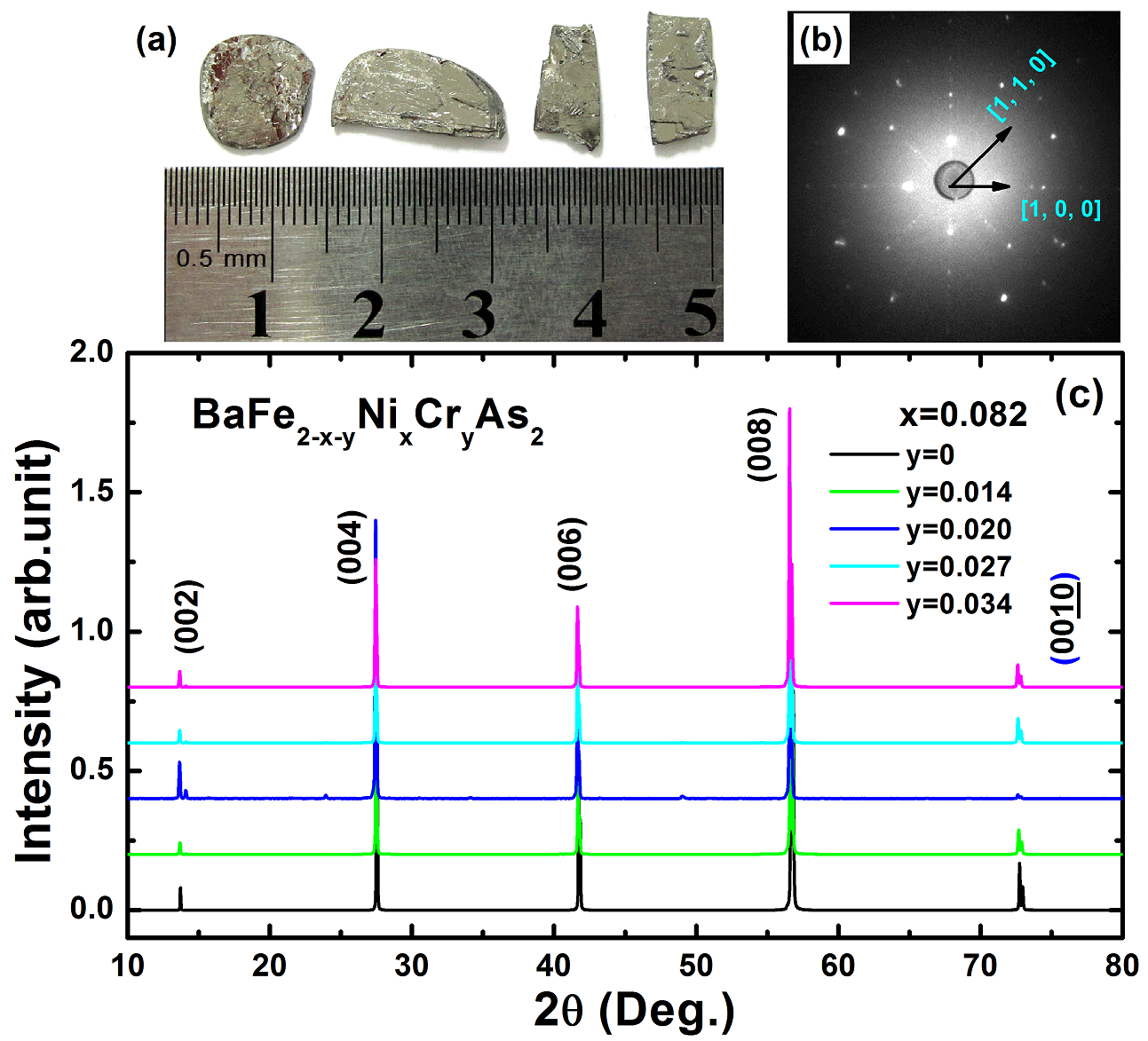}
\caption{
(a) Picture of as-grown BaFe$_{2-x-y}$Ni$_{x}$Cr$_{y}$As$_{2}$ single crystals.
(b) Laue reflection pattern of our crystal.
(c) Typical patterns of X-ray diffraction on the single crystal BaFe$_{2-x-y}$Ni$_{x}$Cr$_{y}$As$_{2}$ with $x=$0.082 and $y=$ 0, 0.014, 0.020, 0.027, 0.034.
 For clarity, the intensity is normalized to [0, 1] and shift upward in 0.2 for each compound.
}
\end{figure}

Single crystals of BaFe$_{2-x-y}$Ni$_{x}$Cr$_{y}$As$_{2}$ were grown by self-flux method similar to BaFe$_{2-x}$Ni$_{x}$As$_{2}$ and Ba$_{1-x}$K$_{x}$Fe$_2$As$_2$ \cite{YCChen,HQLuo2}.
The precursors FeAs, NiAs and CrAs were prepared by solid-state reaction between Fe/Ni/Cr powder and ground As chips at 700 - 900 $^\circ$C.  In order to obtain a homogeneous mixture for the flux, all three precursors were mixed in the appropriate proportions (see Table 1) and heated at 900 - 950 $^\circ$C for 20 hrs, this procedure was repeated for three times. After this, the mixture of Ba pieces and FeNiCrAs flux (in the ratio of 1 : 5) was loaded into an aluminum oxide crucible and sealed in an evacuated quartz tube. The sealed quartz tube was heated up to 950 $^\circ$C in a box furnace, holding for more than 10 hours to completely melt Ba.
The system was then heated to 1180 $^\circ$C in 12 hours and held for 10 hours to melt flux, following by cooling down to 1050 $^\circ$C at a rate of 5 $^\circ$C/h and finally quenched to room temperature. In-situ annealing with flux was tried at 800 $^\circ$C for 20 hrs, which did not significantly affect on the superconductivity.

The crystallinity was examined by a \emph{Photonic Sciences} Laue camera and a Mac-Science MXP18A-HF equipment for x-ray diffraction. The superconducting properties the samples were determined by the DC magnetization measurement using a \emph{Quantum Design} SQUID Magnetic Property Measurement System (\emph{MPMS}). To minimize the effect from geometry and demagnetization factor, we applied a small DC field ($H=$ 10 or 20 Oe) parallel to $ab-$plane and measured using a zero-field cooled protocol. Resistance measurements were carried out on a \emph{Quantum Design} Physical Property Measurement System (\emph{PPMS}) by the standard four-probe method.  The measured crystals were cut into rectangluar shapes $3\times 1\times0.1 $ mm$^3$ with the $c$-axis being the smallest dimension. Four Ohmic contacts with low resistance (less than 1 $ \Omega$) on $ab$-plane were made by silver epoxy.  In order to reduce noises, a large current $I=5$ mA and slow sweeping rate of temperature (2 K/min) were applied.

\section{Result and discussion}

Large single crystals sized up to 15 mm can be cleaved from the as-grown ingot, as shown in Fig. 2 (a). To characterize the quality of our samples, we have performed x-ray diffraction and Laue reflection with incident beam parallels to $c-$aixs. Figure 2 (b) shows the clear pattern with four-fold spots from Laue reflection, where the arrows mark the [1, 0, 0] and [1, 1, 0] directions. Figure 2(c) shows the x-ray diffraction patterns for the sample with $x=$0.082 and $y=$ 0, 0.014, 0.020, 0.027, 0.034 measured at room temperature. All sharp peaks with even index indicate high $c$-axis orientation and crystalline quality of our samples. Since Cr substitution is within Fe-As plane, the $c-$axis does not have significant change for different Cr concentrations, resulting in similar sets of x-ray diffraction data.

The chemical compositions of our crystals are determined by the inductively coupled plasma (ICP) analysis. By
assuming As positions are fully occupied at 2, we can estimate the ratio of Ba : Fe : Ni : Cr as summarized in Tabel. 1. The large deviation of Ba composition away from 1 in several samples is due to residual flux mixed in the samples, which is very common in the self-flux method and does not affect the calculated result of the Ni and Cr concentration. Statistically, the segregation coefficient, namely the ratio between real composition and nominal concentration, $K= C_s/C_l$ for Ni and Cr is about 0.82 $\pm$ 0.10 and 0.68 $\pm$ 0.11, respectively, which is consistent with previous reports on the electron doped Ba-122 system \cite{NNi,YCChen}. In the following discussions of this paper, we use the real composition of Ni doping $x$ and Cr doping $y$ to label all compounds.

\begin{table}

Table 1. Real composition of BaFe$_{2-x-y}$Ni$_{x}$Cr$_{y}$As$_{2}$  single crystals from ICP analysis.\\
\begin{center}
\begin{tabular}{|c|c|c|c|c|c|c|c|}
 \hline \hline
Nominal y/x  & Ba    & Fe    & Ni    & Cr     & As & $K$(Ni) & $K$(Cr)\\
  \hline
0.006/0.06   & 0.76  & 1.92  & 0.06  & 0.004  & 2  & 1.00  & 0.67\\
0.012/0.06   & 0.98  & 1.90  & 0.04  & 0.006  & 2  & 0.67  & 0.50\\
0.008/0.08   & 0.96  & 1.92  & 0.06  & 0.006  & 2  & 0.75  & 0.75\\
0.016/0.08   & 0.94  & 1.90  & 0.06  & 0.012  & 2  & 0.75  & 0.75\\
0.024/0.08   & 0.90  & 1.90  & 0.06  & 0.016  & 2  & 0.75  & 0.67\\
0.01/0.1     & 0.98  & 1.90  & 0.08  & 0.006  & 2  & 0.80  & 0.60\\
0.02/0.1     & 0.98  & 1.88  & 0.08  & 0.014  & 2  & 0.80  & 0.70\\
0.03/0.1     & 1.00  & 1.90  & 0.08  & 0.02   & 2  & 0.80  & 0.67\\
0.04/0.1     & 0.96  & 1.86  & 0.08  & 0.02   & 2  & 0.80  & 0.50\\
0.05/0.1     & 0.98  & 1.86  & 0.08  & 0.04   & 2  & 0.80  & 0.80\\
0.024/0.12   & 0.52  & 1.84  & 0.12  & 0.02   & 2  & 1.00  & 0.83\\
0.036/0.12   & 0.68  & 1.84  & 0.12  & 0.02   & 2  & 1.00  & 0.56\\
0.048/0.12   & 0.84  & 1.84  & 0.10  & 0.04   & 2  & 0.83  & 0.83\\
0.015/0.15   & 0.88  & 1.84  & 0.12  & 0.01   & 2  & 0.80  & 0.67\\
0.03/0.15    & 0.84  & 1.84  & 0.12  & 0.02   & 2  & 0.80  & 0.67\\

\hline \hline
\end{tabular}
\end{center}\label{tab:tableI}
\end{table}

Figure 3 shows the DC susceptibility data for the superconducting samples, where the magnetic signals are normalized by the susceptibility $\chi_0$ at based temperature (2 K or 5 K).
Since the Cr atom has two electrons less than Fe, one would expect hole doping effects in Cr substituted samples \cite{ASSefat2}, contrasting
to the electron doping effect in Ni substituted samples \cite{NNi}. If this is indeed the case,
adding equivalent amount of Cr as Ni to the electron-doped BaFe$_{2-x}$Ni$_{x}$As$_{2}$ should bring the system back to the undoped AF state.
We find that this is clearly not the case.  Instead, superconductivity can be easily suppressed by a small amount of Cr doping, suggesting that the Cr doping acts as an impurity, similar to the cases in Ni or Co doped Ba$_{0.5}$K$_{0.5}$Fe$_2$As$_2$ \cite{JLi2}. In the optimally doped BaFe$_{2-x}$Ni$_{x}$As$_{2}$ with $x=0.082$, the maximum $T_c$ is about 20.1 K in the Cr free sample.  Substituting 0.35 \% Cr with $y=$ 0.007 in BaFe$_{2-x-y}$Ni$_{x}$Cr$_{y}$As$_{2}$, $T_c$ quickly decreases to 15.2 K. No superconductivity above 0.05 K is found in the compounds with $x=$ 0.082 and $y=$ 0.027, indicating that 1.5 \% Cr can suppress optimal superconductivity in BaFe$_{2-x}$Ni$_{x}$As$_{2}$ system.
Such large $T_c$ suppression rate (about 13 K/\%) is similar to the impurity effect to superconductivity in
the hole-doped 122-type iron-based superconductors \cite{PCheng1,JLi1,JLi2,JLi3,PCheng2}.
Surprisingly, the superconducting transition width, defined as the temperature difference between 10 \% and 90 \% of $\chi_0$, is within 2 K in the entire Cr doping range, suggesting high quality of superconductivity in our samples.

\begin{figure}[t]
\center\includegraphics[scale=.5]{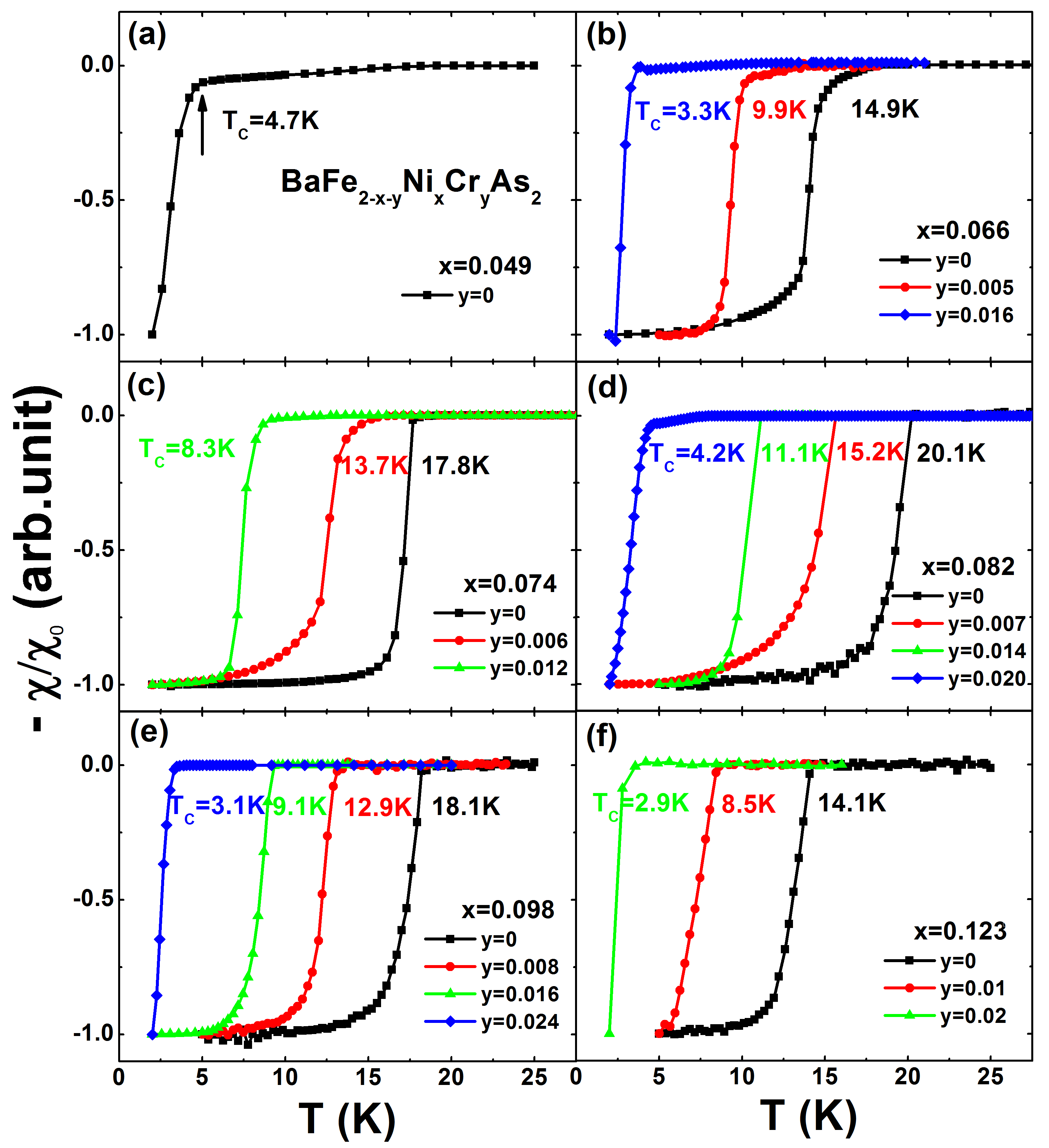}
\caption{
Magnetization and $T_C$ for all superconducting compounds. All signals are normalized by the data $\chi_0$ at based temperature (2 K or 5 K).
}
\end{figure}

\begin{figure}[t]
\center\includegraphics[scale=.5]{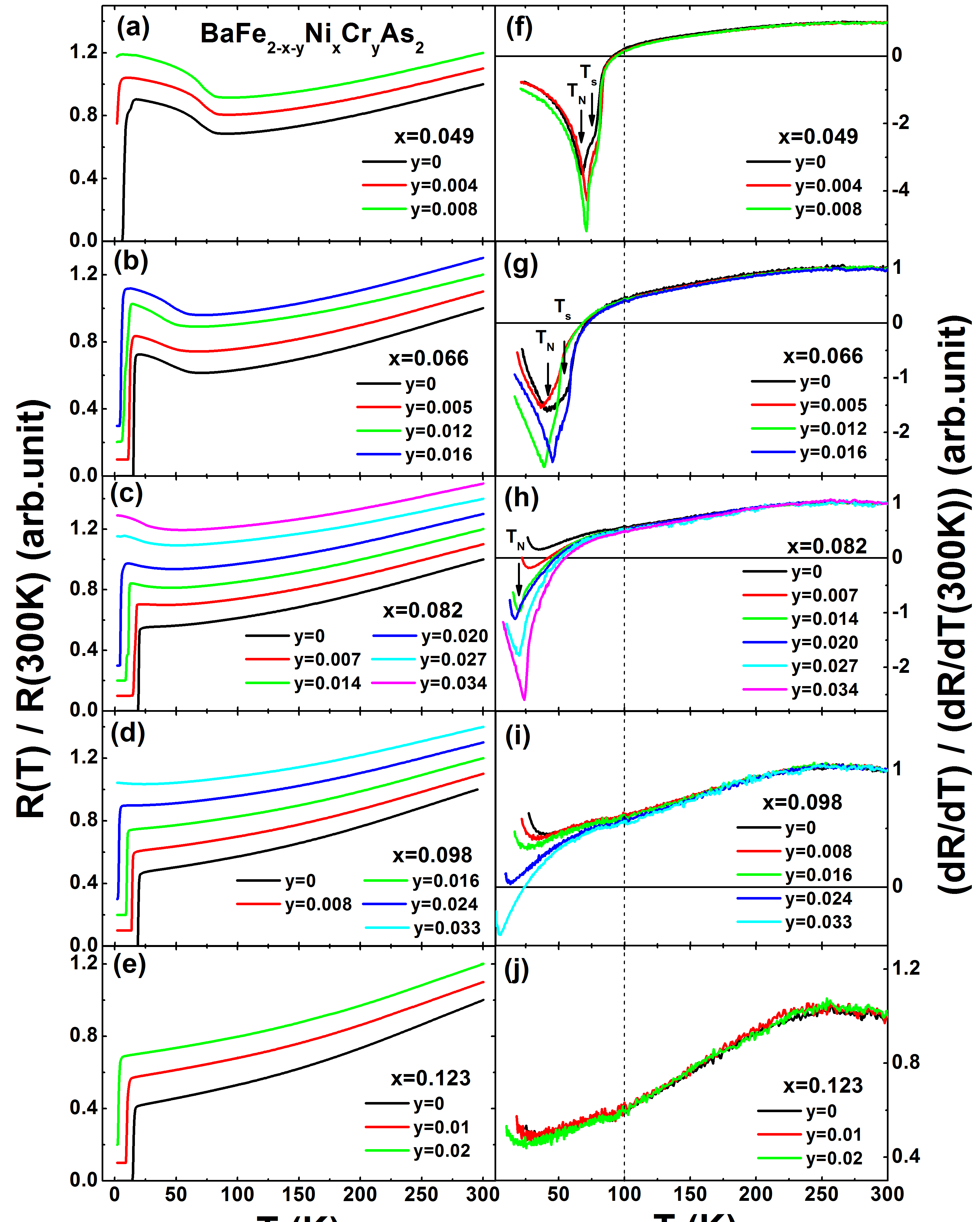}
\caption{
(a)- (e) In-plane resistance up to 300 K for BaFe$_{2-x-y}$Ni$_{x}$Cr$_{y}$As$_{2}$. All data are normalized by
the room temperature resistivity $R$ (300 K), to remove uncertainty
in estimates of the absolute value due to geometric
factors. Each data set is offset vertically by 0.1 for
clarity.
(f) - (j) First order of derivative for the resistance. All curves are also normalized by the data at 300 K, where the arrows show antiferromagnetic transition temperature $T_N$ and structure transition temperature $T_s$.
}
\end{figure}

\begin{figure}[t]
\center\includegraphics[scale=.5]{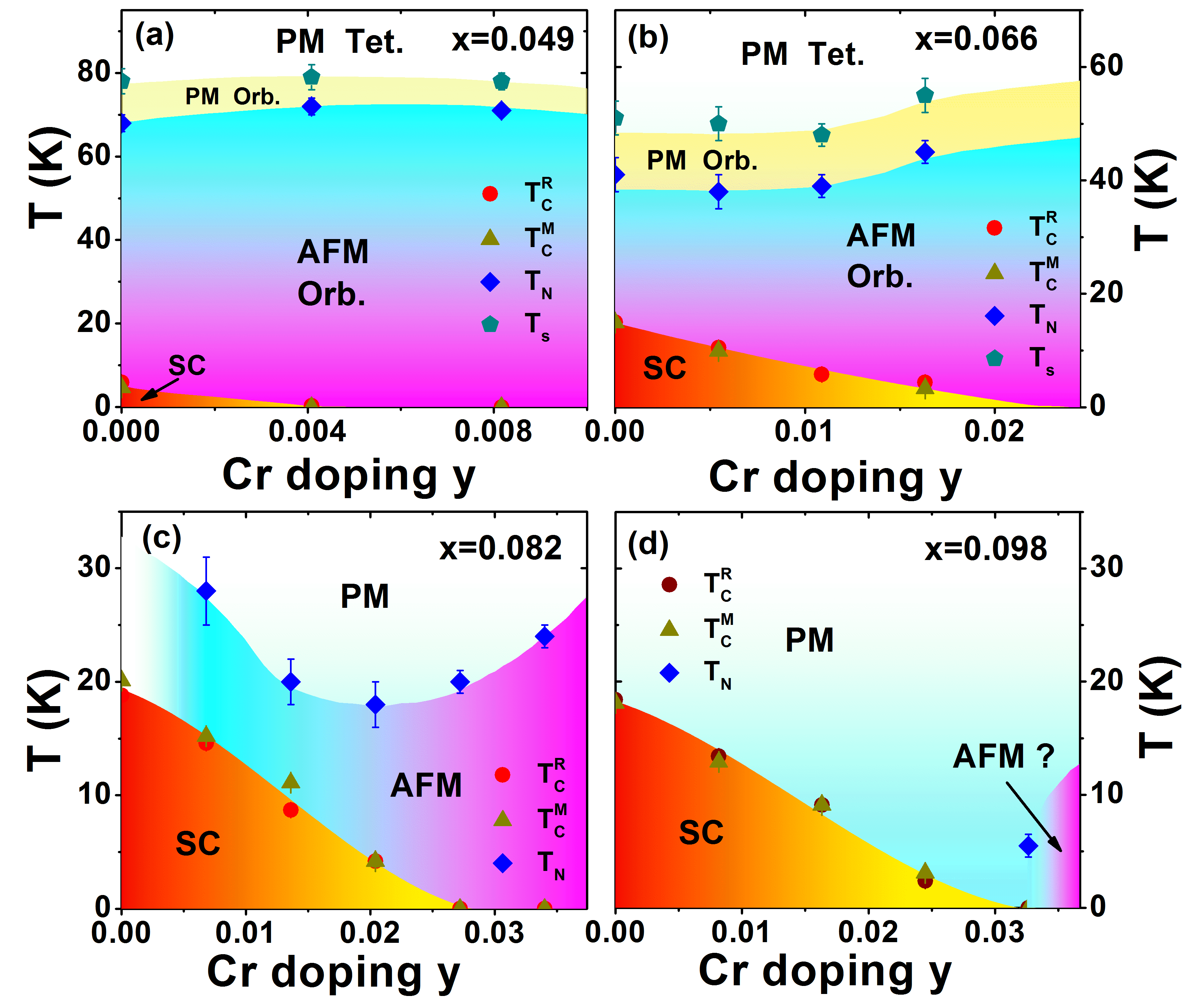}
\caption{
(a)Electronic phase diagram of
BaFe$_{2-x-y}$Ni$_{x}$Cr$_{y}$As$_{2}$ for Ni doping $x=$ 0.049, 0.066, 0.082 and 0.098 under different concentration of Cr impurity, where $T_c^R$ and $T_c^M$ mark the superconducting transition temperature determined by resistance and magnetization, respectively.
}
\end{figure}

To further investigate the normal state behavior in the Cr impurity substituted BaFe$_{2-x-y}$Ni$_{x}$Cr$_{y}$As$_{2}$ single crystals, we have performed systematic resistance measurements on all the compounds up to 300 K.
Figures 3(a)-(e) show temperature dependence of the resistivity data normalized by the room temperature resistivity $R$ (300 K)
to remove the uncertainty
in estimating the absolute value due to geometric factors.
The systematic decrease of $T_c$ with increasing Cr doping confirms the magnetization measurements.
Moreover, the temperature dependence of in-plane resistance are almost identical for temperatures above 100 K in the same batch of Ni doping $x$. This is clearly seen in the first derivative of the resistance data [Figs. 4 (f) - (j)],
where all the normalized $dR/dT$ curves overlap above 100 K for the same Ni concentration.
This suggests that the normal state resistivity behavior is not much affected by the doped Cr impurity.
If the temperature dependence of the resistivity suggests the presence of a quantum critical point,
one should observe linear dependence of resistance or flat temperature dependent region in $dR/dT$.
Inspection of Figs. 4(f)-(j) reveals $R \sim T^2$ behavior in the electron overdoped regime typical of Fermi liquid.

The temperature dependence of the resistivity below 100 K reveals a slight upturn consistent with the Cr impurity scattering.
From the weak anomalies in the temperature dependence of the resistivity \cite{YCChen,JHChu}, we
can identify the AF and tetragonal to orthorhombic structural transitions.
The two dips in $dR/dT$ correspond to $T_N$ and $T_s$ marked as arrows in Fig. 4 (f)-(h).
We note that the small upturn in $dR/dT$ at low temperature comes from the rapid decreasing of $R$ due to the superconducting transition at $T_c$. A sign change in $dR/dT$ is necessary for well-defined $T_N$ and $T_s$.
From the transport data in Figs. 3 and 4, we sketch the phase diagram for the samples with Ni dopings $x=$ 0.049, 0.066, 0.082 and 0.098 in Fig. 5. For the underdoped compounds with $x=$ 0.049 and 0.066, $T_N$ and $T_s$ are essentially independent of the Cr substitution.
By increasing Cr impurity into the optimal doped sample $x=$ 0.082, $T_N$ drops from 30 K to 20 K, and then increases after the suppression of superconductivity. There may be a weak AF order with $T_N=$ 5.5 K in the compound with $x=$ 0.098 and $y=$ 0.033 [Fig. 4(i)], suggesting the competing states between AF order and superconductivity in this system.
Neutron scattering experiments in these materials are currently underway to confirm this conclusion.  We also note
that high doping of Cr into the parent compound BaFe$_2$As$_2$ could also suppress the AF order without change of ordered moments \cite{ASSefat2,KMarty}. Similar to Ba(Fe$_{1-x}$Mn$_x$)$_2$As$_2$ \cite{YSingh,APandey}, large Cr doping ($> 30 \%$) changes the spin structure of the system from C-type to G-type \cite{KMarty}.  Since the Cr concentrations in our BaFe$_{2-x-y}$Ni$_{x}$Cr$_{y}$As$_{2}$ samples are less than 2 \%, we believe that the magnetic structure in the Cr-doped samples are the same as that of the undoped compounds. Moreover, in our recent collaborative measurements by Angle-Resolved-Photoemission-Spectroscopy (ARPES) and thermal conductivity, we have found the band structure and Fermi surfaces are barely changed upon doping Cr into BaFe$_{2-x}$Ni$_{x}$As$_{2}$ with same Ni concentration, except for weak localization effects of charge carriers and suppression of $T_c$ \cite{RZhang}. These are consistent with our resistivity results and further confirm the impurity effects from Cr dopings.

\section{Summary}

In summary, we have carried out systematic transport and magnetic measurements on Cr-doped
BaFe$_{2-x-y}$Ni$_{x}$Cr$_{y}$As$_{2}$ single crystals.  We find that Cr-doping is very efficient in suppressing superconductivity and
can be regarded as an impurity instead of hole-doping. The AF ordering transition and structure transition in the underdoped samples are hardly affected by the small amount of Cr doping, while the antifferromgantic correlations may be restored in optimally and overdoped samples when superconductivity is eliminated by Cr doping.

\section{Acknowledgments}

This work is supported by the National Science Foundation
of China (No. 11374011), the Ministry of Science and Technology of China
(973 projects: 2011CBA00110 and 2012CB821400), and The Strategic Priority Research Program
(B) of the Chinese Academy of Sciences (Grant No.
XDB07020300).The work at Rice is supported by NSF DMR-1362219 and the Robert A.
Welch Foundation Grant No. C-1839. The authors thank the
fruitful discussion with Tao Xiang, and the great
help on the crystal characterization from Lihong Yang, Jun Zhu, Cong Ren and Xingjiang Zhou.

\end{document}